\documentstyle[prc,preprint,aps]{revtex}
\begin{document}
\title{CHIRAL BACKGROUND FOR THE TWO PION EXCHANGE NUCLEAR POTENTIAL: A
PARAMETRIZED VERSION}
\author{C.A. da Rocha}
\address{Instituto de F\'{\i}sica Te\'orica,
Universidade Estadual Paulista \\
R. Pamplona, 145 - 01405-900 - S\~ao Paulo, SP, Brazil,
e-mail: carocha@vax.ift.unesp.br}
\author{M.R. Robilotta}
\address{Instituto de F\'{\i}sica, Universidade de S\~ao Paulo \\
C.P. 20516, 01452-990, S\~ao Paulo, SP, Brazil,
e-mail: robilotta@if.usp.br}
\date{\today}
\maketitle
\begin{abstract}
We argue that the minimal chiral background for the two-pion exchange
nucleon-nucleon interaction has nowadays a rather firm conceptual basis,
which entitles it to become a standard ingredient of any modern potential.
In order to facilitate applications, we present a parametrized version of a
configuration space potential derived previously. We then use it to assess
the phenomenological contents of some existing NN potentials.
\end{abstract}
\pacs{12.39Fe - 21.30.+y - 13.75.Cs}

\newpage
\section{Introduction}

Nowadays there are several nucleon-nucleon potentials which reproduce
experimental information with high precision
\cite{TS73,TRS75,Lac+80,MHE87,WSA84,Nij78}. In all of them, the long and medium
range interactions are associated with the exchanges of one and two pions,
whereas there is a wide variation in the way short distance effects are
treated. This region is theoretically uncertain and good fits to the data
usually require many free parameters. The availability of alternative models,
although welcome, points out to the need of criteria for discriminating among
the various potentials.

All potentials become progressively less reliable as one goes from large to
short distances. The pionic tail of the interaction is quite well established
and its only source of uncertainty is the value of the pion-nucleon coupling
constant. It determines many observables of the deuteron and large angular
momentum scattering \cite{ER82,BER89,BR92,BR95}.
As far as the next layer is concerned, one finds some consensus in the
literature regarding the importance of delta intermediate states and pion-pion
correlations. Nevertheless, the treatment of details of these interactions is
not uniform, a fact reflected in the existence of different profile functions
for a given component of the potential. It is in this framework that chiral
symmetry may prove to be useful and provide guide-lines for a proper
assessment of the various existing potentials.

Chiral symmetry corresponds to the idea that strong interactions are
approximately invariant under transformations of the group $SU(2)\times SU(2)$.
This symmetry was developed in the sixties, in the framework of hadron physics,
but the reasons behind its successes became clear only after the formulation
of QCD, the theory decribing the basic interactions of quarks and gluons.
In QCD the quark masses of the low-lying $SU(2)$ multiplet
are very small and the Lagrangian becomes chiral invariant when they are
neglected. QCD predictions can be directly tested only at high energies, where
perturbation makes sense. At low energies, on the other hand, its non-Abelian
structure makes calculations very difficult, and one is forced to resort to
effective hadronic theories. In order to remain as close as possible to the
fundamental level, the hadronic Lagrangian must have the same symmetries as
QCD, even the approximate ones, such as chiral symmetry.

One of the major successes of chiral symmetry at hadron level were the
predictions made for low-energy $\pi$N scattering. Early chiral calculations of
this process considered just minimal systems, containing only pions and
nucleons \cite{Wei66,Tom67}. Motivated by phenomenology, this model was
extended and one has learned that chiral amplitudes at tree level, including
nucleon and delta poles, rho exchanges and corrections associated with $\pi$N
sigma term do reproduce satisfactorily subthreshold and scattering data up to
300 MeV.

Chiral symmetry at hadron level may be implemented by means of either linear or
non-linear Lagrangians. The fact that no serious candidate has been found for
the $\sigma$ meson favours the latter type of approach. There are two forms of
non-linear Lagrangians which are especially suited for the $\pi$N system. One
of them is based on a pseudoscalar (PS) $\pi$N coupling supplemented by a
scalar (S) interaction, equivalent to the exchange of
an infinitely massive $\sigma$ meson, and
denoted as $PS+S$ scheme (Eq.~\ref{Eq.1}). The other one employs a pseudovector
(PV) $\pi$N interaction and a vector (V) term, which could represent the
exchange of an infinitely massive $\rho$ meson, constituting the $PV+V$ scheme
(Eq.~\ref{Eq.2}).
\begin{equation}
{\cal L}_{PS+S} = \cdots - g \overline{\psi}\,
  \left(\sqrt{f_{\pi}^2 - \phi^2} + i \bbox{\tau} \cdot \bbox{\phi}
  \gamma_5\right) \psi\quad ,
\label{Eq.1}
\end{equation}
\begin{equation}
{\cal L}_{PV+V} = \cdots - \frac{1}{4 f_{\pi}^2}\,
   \overline{\psi}\,\gamma^\mu\,\bbox{\tau}\,\psi\cdot
   \bbox{\phi} \times \partial_\mu\,\bbox{\phi}
   +\frac{g}{2m}\,\overline{\psi}\,\gamma^{\mu}
   \gamma_5\, \bbox{\tau} \,\psi\cdot\partial_\mu\,
  \bbox{\phi} + \cdots\;.
\label{Eq.2}
\end{equation}

\noindent
Both approaches yield the very same amplitude for the $\pi$N scattering when
the axial coupling constant $g_A$ is equal to 1. The extention to the case
$g_A\not =1$ can be done quite naturally in the $PV+V$ approach, and a
little less so in the $PS+S$ case. The fact that physical results should be
independent of the representation used to implement chiral symmetry was
discussed in very general terms by Coleman, Wess, and Zumino \cite{CWZ69}. In
the case of $\pi$N interactions, this point was emphasized by Coon and Friar
\cite{CF86}.

Chiral symmetry has very little to say about the one pion exchange NN potential
since, in the $PS+S$ approach, predictions for this component do not depend
on the scalar interaction. Moreover, $PS$ and $PV$ couplings are equivalent.
The intermediate  range
part of the interaction, on the other hand, is due to the exchange of two pions
and closely related to an off-shell $\pi$N amplitude. Thus, as pointed out by
Brown and Durso in 1971 \cite{BD71}, chiral symmetry is quite relevant in this
region. In the first calculations of this component of the potential, chiral
symmetry was included disguised in the form of soft-pion theorems
\cite{BD71,CDR72}. Nowadays one has realized that the use of effective field
theories makes the implementation of chiral symmetry into hadronic systems
much easier.

In the early seventies, approaches to the intermediate range NN interactions
based on field
theory disconsidered chiral symmetry. Nevertheless, they were very important
in setting the method to tackle the problem. In 1970, Partovi and Lomon
\cite{PL70} performed a paradigmatic calculation based on Feynman diagrams,
using just $PS$ couplings without chiral counterterms. A little later, Partovi
and Lomon \cite{PL72} included the $\sigma$ meson in their calculation, used
chiral symmetry to determine the $\sigma$N coupling constant, but did not group
diagrams into chiral families, producing a potential with an uneven number of
loops. About ten years later a rather comprehensive study was performed by
Zuilhof and Tjon \cite{ZT82}, who compared direct and crossed box diagrams
obtained using both PS and PV couplings, without paying attention to chiral
symmetry.

The first modern field theoretical calculation of the two-pion exchange
nucleon-nucleon potential ($\pi\pi$E-NNP) with strong emphasis on chiral
symmetry was produced by Ord\'o\~nez and van Kolck \cite{OVK92}. They applied
the conceptual pertubative framework developed by Weinberg \cite{Wei90} in a
systematic discussion of corrections to pure OPE interactions,
associated with form factor effects, exchange of two pions, and three body
forces. The problem was formulated using $PV$ coupling for the pion, and
expressions were given in momentum space.

In 1992, Celenza, Pantziris and Shakin \cite{CPS92} studied the influence
of chiral symmetry on the NN interaction at one-loop order. These authors
considered both $PS$ and $PV$ pion-nucleon couplings in order to derive the
qualitative features of the scalar-isoscalar component of the interaction.
They employed tensor decompositions of diagrams in momentum space, and
one of their main conclusions was that the irreducible isoscalar interaction
is attractive.

In a series of independent works, other authors also addressed to the same
problem, namely the determination of the role of chiral symmetry in the
two-pion exchange NN interaction. A momentum space evaluation of the
scalar-isoscalar component was produced by Birse \cite{Bir94}, in the framework
of the linear $\sigma$-model. Friar and Coon \cite{FC94}, used a more general
Lagrangian in order to obtain a momentum space potential for infinitely massive
nucleons. Finally, in another calculation, we \cite{RR94} used non-linear
Lagrangians based on covariant derivatives and obtained a non-relativistic
potential in coordinate space.

All these calculations \cite{OVK92,CPS92,Bir94,FC94,RR94} are based on chiral
symmetry and should, in spite of technical differences, yield the same
predictions for the two-pion exchange potential. However, inspection of the
conclusions presented by the various authors shows that this is not the case.
Discrepancies may be traced back to two different sources, namely the methods
employed for regularizing divergent integrals and the procedures used to
subtract the iterated OPEP. We here consider only the latter problem. It is
well known that the meaning of an iteration is defined by the dynamical
equation adopted in a given problem. Thus, the use of the relativistic
Bethe-Salpeter equation \cite{SB51} or the non-relativistic reductions due to
Brueckner and Watson \cite{BW53} or Blankenbecler-Sugar \cite{BS66} correspond
to
different forms for the potential. An explicit discussion of the
non-relativistic case can be found in Ref.~\cite{FC94}, where it is shown that
physics is independent of the particular approach employed.

Another important feature of the problem is that physical results in a
consistent calculation should also not depend
 the particular representation used
in the implementation of chiral symmetry \cite{CWZ69,CF86}. In our calculation
\cite{RR94}, we have shown explicitly that both $PS+S$ and $PV+V$ Lagrangians
correspond to the same non-relativistic potential. Of course, there is no
reason to expect that things would be different in the case of the
Bethe-Salpeter equation. However, we believe that the explicit demonstration
of this property in the relativistic framework may prove to be non-trivial.
The basic idea of the Bethe-Salpeter equation is to decompose the full
amplitude into two-particle propagators and irreducible kernels. In the $PS+S$
case, this kind of decomposition poses no special problem. The $PV+V$ case, on
the other hand, is tricky, since its box diagram contains contributions in
which the pole of the nucleon propagator is canceled and hence must be
interpreted with great care. Cancellations of nucleon poles are, in fact,
already present in the intermediate $\pi$N amplitude that determines the
$\pi\pi$E-NNP \cite{RR94,Pup94}. Therefore, in the present state of the art,
we believe that non-relativistic potentials provide a much safer ground for the
study of chiral symmetry.

The minimal chiral model for the $\pi\pi$E-NNP, involving only pions and
nucleons, has nowadays very solid conceptual foundations. It has attained the
same status as the OPEP in the late sixties and one may expect it to become a
necessary ingredient of any modern NN potential. On the other hand, this
minimal chiral model fails to reproduce experimental information in the case
of the intermediate $\pi$N amplitude incorporated into the NN interaction.
Thus, in the $\pi\pi$E-NNP, the minimal model
corresponds to a contribution which is necessary, but not sufficient to
describe reality. This requirement demands the inclusion of other fields
and processes in the model. An extension, in the framework of chiral
symmetry, has been recently considered by Ord\'o\~nez, Ray and van Kolck
\cite{ORK94}, with promissing results.

In all calculations of the chiral background to the $\pi\pi$E-NNP, results are
given in form of cumbersome integrals, which must be evaluated numerically. In
order to facilitate applications, in Sec. II of this work we present a
parametrized version of our configuration space potential, valid up to 0.5 fm
and completely free of cut off parameters. The knowledge of the minimal chiral
$\pi\pi$E-NNP may be used to assess the phenomenological content of a given
potential. This information can be obtained by subtracting the OPEP from it and
then comparing the remainder with the chiral two pion
exchange background. In Sec. III, we consider
the dTRS \cite{TRS75}, Paris \cite{Lac+80}, and Argonne \cite{WSA84} potentials
in order to produce an instance of this kind of study.

\section{Parametrization}

Our calculation of the $\pi\pi$E-NNP is based on the Blankenbecler-Sugar
reduction of the Bethe-Salpeter equation. In the $PS+S$ scheme, its dynamical
content is associated with the five diagrams displayed in Fig.~\ref{Fig.1}.
Therefore we label the corresponding individual contributions by
${(\!)}\;,\triangle\;,{\Join}\mbox{ and }\Box$, where the last one also
includes
the subtraction of the iterated OPEP. It has the general form\footnote{See
Eq.(59) of Ref.~\cite{RR94}.}

\begin{eqnarray}
V(r)&=&\left[\left(V_{(\!)}^C+V_\triangle^C\right)+\hat{O}_{LS}
\left(V_{(\!)}^{LS}+V_\triangle^{LS}\right)\right]+ \nonumber \\
&&\left(3+2 \bbox{\tau}^{(1)}\cdot \bbox{\tau}^{(2)}\right)
\left[V^C_{\Join}+\hat{O}_{SS} V^{SS}_{\Join}+\hat{O}_{LS} V^{LS}_{\Join}
+\hat{O}_T V^{T}_{\Join}\right] + \nonumber \\
&&\left(3 - 2\bbox{\tau}^{(1)}\cdot\bbox{\tau}^{(2)}\right)
\left[V_\Box^C+\hat{O}_{SS}V_\Box^{SS}+\hat{O}_{LS}V_\Box^{LS}+\hat{O}_T
V_\Box^T\right]
\label{Eq9}
\end{eqnarray}
\noindent
where the spin operators are given by $\hat{O}_{SS}=\bbox{\sigma}^{(1)}\cdot
\bbox{\sigma}^{(2)}$, $\hat{O}_{LS}={\bf
L}\cdot\case{1}/{2}\{\bbox{\sigma}^{(1)}+
\bbox{\sigma}^{(2)}\}$, and $\hat{O}_T=3\bbox{\sigma}^{(1)}\cdot\hat{\bf r}
\;\;\bbox{\sigma}^{(2)}\cdot\hat{\bf r}-\bbox{\sigma}^{(1)}\cdot
\bbox{\sigma}^{(2)}$, whereas $\bbox{\sigma}^{(i)}$ and $\bbox{\tau}^{(i)}$
represent spin and isospin matrices for nucleon $(i)$.

In the case of the bubble diagram, the leading contribution to the asymptotic
potential can be calculated analytically, as shown in appendix. This results
sets the pattern for the parametrization of the other components of the force.

Our numerical expressions represent the various components of the potential in
MeV, and are given in terms of the adimensional variable $x\equiv\mu r$, where
$\mu$ is the pion mass. We keep the $\pi$N coupling constant $g$ as a free
parameter  and adopt the values $\mu=137.29$ MeV and $m=938.92$ MeV for the
pion and nucleon masses respectively.

In general, the parametrized expressions reproduce quite well the numerical
results of Ref.~\cite{RR94}, except for a few cases and regions where the
discrepancies become of the order of 0.25\%. Our results are listed below.

\subsection{Central Potential}

The profile function for the central potential has the following common
multiplicative expression
\begin{equation}
F_c(x)=\left(\frac{g\mu}{2m}\right)^4\;\frac{e^{-2x}}{x^2\sqrt{x}}\;.
\end{equation}

The parametrization of each diagram gives:
\begin{eqnarray}
V^C_{(\!)}(x)&=&F_c(x)\left\{-275.364-\frac{51.0923}{x}+\frac{6.54068}{x^2}-
\frac{1.26190}{x^3}+\frac{0.130706}{x^4}\right\}
\label{vc01} \\ [0.3cm]
V^C_{\triangle}(x)&=&F_c(x)\left\{343.558-\frac{14.0446}{x}
\right. \nonumber \\
&&\left.+\left(135.249+14.6514\cdot x+6.43825\cdot x^2\right)\cdot
e^{-0.397835\cdot x}\right\}
\label{vc02}
\\ [0.3cm]
V^C_{\Join}(x)&=&\frac{V^C_{\triangle}(x)}{12}+F_c(x)\left\{
-\frac{265.304}{\sqrt{x}}+\frac{518.531}{x}\right.\nonumber \\
&&\left.-\frac{577.210}{x\sqrt{x}}+\frac{378.004}{x^2}-
\frac{133.374}{x^2\sqrt{x}}+\frac{19.5061}{x^3}\right\}
\label{vc12}
\\ [0.3cm]
V^C_{\Box}(x)&=&F_c(x)\left\{-25.9987+\frac{8.33777}{x}-\frac{0.870724}{x^2}
\right\}\cdot e^{-[0.101214\cdot x+0.00123687\cdot x^2]}
\label{vc09}
\end{eqnarray}

\subsection{Spin-spin potential}

The multiplicative factor for the spin-spin potential is the same as that of
the central potential, and receives contributions from the box and crossed
diagrams only, which are given by
\begin{eqnarray}
V^{SS}_{\Join}(x)&=&F_c(x)\left\{0.408084+\frac{1.05042}{x}+\frac{0.421043}
{x^2}-\frac{0.0284309}{x^3}-0.215829\;\cdot e^{-1.2344\cdot x}\right\}
\label{vss12} \\ [0.3cm]
V^{SS}_{\Box}(x)&=&F_c(x)\left\{0.399845+\frac{1.07191}{x}+\frac{0.216302}{x^2}
-\frac{0.0306271}{x^3}+0.0371333\,x\cdot e^{-0.16808\cdot x}\right\}
\label{vss09}
\end{eqnarray}

\subsection{Spin-orbit Potential}

The spin-orbit multiplicative function is
\begin{equation}
F_{LS}(x)=\left(\frac{g\mu}{2m}\right)^4\left(1+\frac{1}{2x}\right)
\;\frac{e^{-2 x}}{x^3\sqrt{x}}\;,
\end{equation}
\noindent
and individual contributions are:
\begin{eqnarray}
V^{LS}_{(\!)}(x)&=&F_{LS}(x)\left\{-5.88744-\frac{5.51078}{x}+\frac{0.994157}
{x^2}-\frac{0.217562}{x^3}+\frac{0.0336067}{x^4}-\frac{0.00244620}{x^5}
\right\}
\label{vls01} \\ [0.3cm]
V^{LS}_{\triangle}(x)&=&F_{LS}(x)\left\{7.34548+\frac{2.15233}{x}-
\frac{0.381025}{x^2}\right. \nonumber \\
&&\left.+\left(7.48798-0.448484\cdot x+0.391431\cdot x^2\right)
\cdot e^{-0.419984\cdot x}\right\}
\label{vls02} \\ [0.3cm]
V^{LS}_{\Join}(x)&=&-\frac{V^{LS}_{\triangle}(x)}{4}+F_{LS}(x)\left\{
\frac{5.67235}{\sqrt{x}}-\frac{10.6417}{x}\right.\nonumber \\
&&\left.+\frac{14.7389}{x\sqrt{x}}-\frac{12.6506}{x^2}+\frac{6.27374}
{x^2\sqrt{x}}-\frac{1.66637}{x^3}+\frac{0.184264}{x^3\sqrt{x}}\right\}
\label{vls12} \\ [0.3cm]
V^{LS}_{\Box}(x)&=&F_{LS}(x)\left\{-1.19527-\frac{1.58089}{x}+\frac{0.319790}
{x^2}-\frac{0.0196461}{x^3}\right\}\cdot\left[1-0.0160259\cdot x\right]^{-1}
\label{vls09}
\end{eqnarray}

\subsection{Tensor Potential}

The commom factor for the tensor potential is
\begin{equation}
F_T(x)=\left(\frac{g\mu}{2m}\right)^4\left(1+\frac{3}{2x}+
\frac{3}{4x^2}\right)\;\frac{e^{-2 x}}{x^2\sqrt{x}}\;.
\end{equation}
\noindent
It receives contributions from the box and crossed diagrams only, which have
the form
\begin{eqnarray}
V^T_{\Join}(x)&=&F_T(x)\left\{-0.204041-\frac{0.510720}{x}+\frac{0.0597556}
{x^2}\right. \nonumber \\
&&\left.+\left(1.32932-0.939553\cdot x+0.706050\cdot x^2\right)
\cdot e^{-2.29686\cdot x}\right\}\label{vt12} \\ [0.3cm]
V^T_{\Box}(x)&=&F_T(x)\left\{-0.246349-\frac{0.521123}{x}+\frac{0.352463}{x^2}
-\frac{0.135028}{x^3}+\frac{0.0303012}{x^4}-\frac{0.00297840}{x^5}
\right\}
\label{vt09}
\end{eqnarray}

\section{Phenomenology}

As discussed in the introduction, one may expect a realistic two-pion exchange
nucleon-nucleon potential to be composed of a minimal chiral background
supplemented by other terms determined by phenomenology. In this section we
estimate crudely the relative weights of these two types of contributions,
by comparing the parametrized expressions of the preceding section with tree
realistic potentials avaliable in the litterature, known as dTRS \cite{TRS75},
Paris \cite{Lac+80}, and Argonne v14 \cite{WSA84}.

In the minimal chiral potential the $\pi$N coupling constant may be kept as a
free parameter, whereas in the realistic versions considered it assumes
slightly different values. Therefore, in order to compare properly the various
results, we subtract the OPEP contribution from each potential and then
divide the remainder by the fourth power of the $\pi$N coupling constant used
in it. We then consider, for various channels, the ratio of this quantity
by that obtained from the parametrization of Sec. II with $g=1$.

The ratios for the ($T=1,\,S=0$) central potential are displayed in
Fig.~\ref{Fig.2}. For distances greater than 7.5 fm, the ratio of the Argonne
potential becomes a constant since, as the minimal chiral background, it
behaves asymptotically as Yukawa functions with double pion mass. In the case
of the dTRS potential, the ratio vanishes because it is relatively short
ranged. The tail of the ratio of the Paris potential, on the other hand, is
rather large and not stable up to 10 fm, probably due to the way it is
parametrized.

The realistic potentials considered here reproduce well NN observables due to
their behaviour in the region below 2.5 fm. Inspecting Fig.~\ref{Fig.2} one
note that, in spite of important differences in this region, the three
potentials display a sort of rough coherence. In all cases the fit of
observables  seem to require contributions with the same sign and about four
times bigger than the chiral background. The case of ($T=1,\,S=1$) central
component is similar. The rough coherence among the realistic potentials is
also present in the ($T=0,\,S=0$) and ($T=0,\,S=1$) central channels, but the
ratios in the region between 1 and 2.5 fm are negative.

In Fig.~\ref{Fig.3} we show the profile function for $V^C$, the spin-isospin
symmetric component of the central potential, which corresponds to the
combination $V^C=V^C_{(\!)}+V^C_\triangle+3V^C_\Box+3V^C_{\Join}$, and is
relevant for symmetric nuclear matter. The chiral background, in the physically
relevant region, is much less attractive than the realistic forces.

In Fig.~\ref{Fig.4} we display the ratio of the ($T=0,\,S=1$) spin-orbit
potentials. In this case there is no coherence among the realistic potentials,
even regarding their signs. Both the tails and medium distance behaviours
of the various ratios differ significantly from the background. The
($T=1,\,S=1$) component behaves similarly.

Finally, the ratios for the $(T=1,\,S=1$) tensor potential are given in
Fig.~\ref{Fig.5}, where is possible to note that the realistic potentials
are roughly coherent below 2.5 fm and about 3 times greater than the
background. In the case of the ($T=0,\,S=1$) component, on the other hand,
this ratio is about 5.

In summary, the crude comparisions made in this section allow one to conclude
that the minimal chiral background for the two-pion exchange nuclear potential
accounts for less than 25\% of the values need in order to reproduce
observations.

\acknowledgments

The work of one of us (C.A.R.) was supported by FAPESP,
Brazilian Agency.

\appendix
\section*{Asymptotic behavior of the bubble diagram}

The asymptotic behaviour of the bubble diagram [Fig.~\ref{Fig.1}(a)] can
be calculated analytically. Its contribution to the potential is given by
\cite{RR94}
\begin{equation}
{V}_{{(\!)}}=i6\,\frac{g^4}{4m^2}\;I^{(1)}\cdot I^{(2)}\cdot J_1 \;,
\label{Eq.A1}
\end{equation}
\noindent
where
\begin{equation}
J_1=\int\frac{d^4\,k}{(2\pi)^4}\frac{1}{(k^2-\mu^2)(k'^{\;2}-\mu^2)}\;,
\end{equation}
\noindent
and  the operator $I^{(1)}\cdot I^{(2)}$ has the following non-relativistic
spin structure
\begin{equation}
I^{(1)}\cdot I^{(2)} = 1 - \frac{\hat{O}_{LS}}{2m^{\,2}} \;,
\label{Eq.C}
\end{equation}
\noindent
$\hat{O}_{LS}$ being the spin-orbit operator.

After the usual integration, the function $J_1$ may be written as
\begin{equation}
J_1(\bbox{\Delta})=\frac{i}{(4\pi)^2}\;(\Lambda^2-\mu^2) \int_0^1d\alpha
\int_0^{1-\alpha}d\beta\frac{1}{\bbox{\Delta}^2\alpha(1-\alpha)+
\mu^2(1-\beta)+ \beta\Lambda^2}\;,
\end{equation}
\noindent
where $\bbox{\Delta}$ is the momentum exchanged in the interaction and
$\Lambda$
is a monopole cut off parameter. The corresponding expression in configuration
space is given by
\begin{equation}
J_1(r)=i\;\frac{\Lambda^2-\mu^2}{(4\pi)^3}\;4\int_0^1d\alpha\int_0^{1-\alpha}
d\beta\frac{1}{(\alpha+\beta)-(\alpha-\beta)^2}\;
\frac{e^{-M_1 r}}{r}\;,
\end{equation}
\noindent
where
\begin{equation}
M_1=2\left[\frac{(1-\alpha-\beta)\Lambda^2+(\alpha+\beta)\mu^2}{
(\alpha+\beta)-(\alpha-\beta)^2}\right]^{1/2}\;.
\end{equation}

In the limit $\Lambda\rightarrow\infty$, the integrand is non-vanishing only
in the neighbourhood of the point $(1-\alpha-\beta)=0$. That means that,
except in the term proportional to $\Lambda^2$, we may set $\beta=1-\alpha$.
With this approximation, one has
\begin{equation}
J_1(r)=i\,4\;\frac{\Lambda^2-\mu^2}{(4\pi)^3}\int_0^1d\alpha\int_0^{1-\alpha}\;
d\beta\;\frac{1}{D^2\;r}\;e^{-2\left[(1-\alpha-\beta)\Lambda^2-\mu^2\right]^
{1/2}\cdot\frac{r}{D}}\;,
\end{equation}
\noindent
where
\begin{equation}
D=1-(2\alpha-1)^2\;.
\end{equation}
\noindent
Defining a new variable $s$ by
\begin{equation}
s^2=(1-\alpha-\beta)\Lambda^2+\mu^2\;,
\end{equation}
\noindent one has
\begin{equation}
J_1(r)=i\frac{8}{(4\pi)^3}\int_0^1d\alpha\;\frac{1}{D^2\;r}\int_\mu^\infty ds
\,s\,e^{-\frac{2sr}{D}}\;.
\end{equation}
\noindent
In deriving this result we used the fact that $\Lambda$ is very large.
Performing the integration and using the variable $x\equiv\mu r$, one obtains
\begin{equation}
J_1(r)=i\frac{4\mu^3}{(4\pi)^3}\int_0^1d\alpha\left[\frac{1}{D\,x^2}+
\frac{1}{2\,x^3}\right]e^{-\frac{2x}{D}}\;.
\end{equation}

For very large values of x, the main contribution to the integral comes from
the region around $\alpha=\case{1}/{2}$, where $D$ is maximum. In order to
explore this property, we use the variable $\theta=2\sqrt{x}(\alpha-
\case{1}/{2})$, expand $D$ around $\theta=0$ and write
\begin{equation}
J_1(x)=i\frac{2\mu^3}{(4\pi)^3}\;\frac{e^{-2x}}{x^2\sqrt{x}}\int_{\sqrt{x}}
^{\sqrt{x}}d\theta\left[1+\frac{1}{2x}\left(1+\theta^2-\frac{3}{2}\theta^4
\right)\right]e^{-\theta^2}\;.
\end{equation}
\noindent
As $x$ is large, the integral can be performed and one has
\begin{equation}
J_1(x)=i\frac{2\mu^3}{(4\pi)^3}\;\sqrt{\pi}\;\left(1+\frac{3}{16x}\right)
\frac{e^{-2x}}{x^2\sqrt{x}}\;.
\end{equation}

The potentials are given by Eq.~\ref{Eq.A1}:
\begin{eqnarray}
V^C_{(\!)}(x)&=&-\left(\frac{g\mu}{2m}\right)^4\frac{e^{-2x}}{x^2\sqrt{x}}
\left[\frac{48\sqrt{\pi}}{(4\pi)^3}\;\frac{m^2}{\mu}\right]\left(1+\frac{3}
{16x}\right)\;; \label{Eq.A} \\
V^{LS}_{(\!)}(x)&=&-\left(\frac{g\mu}{2m}\right)^4\left(1+\frac{1}{2x}\right)
\frac{e^{-2x}}{x^3\sqrt{x}}\left[\frac{48\sqrt{\pi}}{(4\pi)^3}\;\mu\right]
\left(1+\frac{15}{16x}+\frac{21}{64x^2}\right)\;;
\label{Eq.B}
\end{eqnarray}

The numerical value of the terms within square brackets is 275.299 for the
central potential and 5.88607 for the spin-orbit potential, and they should
be compared with those given in Eqs.~(\ref{vc01}) and (\ref{vls01}),
respectively.
This result motivated the structure of the expressions used in the
parametrization of the other components of the potential.

\begin{figure}
\caption{Diagrams contributing to the minimal chiral background for the
$\pi\pi$E-NNP. The vertices with one and two pions may be given by either
PS and S or PV and V couplings \protect\cite{RR94}. The last diagram
represents the subtraction of the iterated OPE from the reducible box diagram.}
\label{Fig.1}
\end{figure}

\begin{figure}
\caption{Ratio between the non-OPEP part of the central ($T=1,\,S=0$) component
of the dTRS \protect\cite{TRS75} (dashed line), Argonne v14
\protect\cite{WSA84} (dot-dashed line) and Paris \protect\cite{Lac+80}
(continuous line) potentials divided by $g^4$ and the minimal chiral
background for the $\pi\pi$E-NNP.}
\label{Fig.2}
\end{figure}

\begin{figure}
\caption{Profile functions for the spin-isospin symmetric component of the
dTRS \protect\cite{TRS75} (dashed line), Argonne v14
\protect\cite{WSA84} (dot-dashed line), Paris \protect\cite{Lac+80}
(continuous line), and the chiral background (dotted line) potentials.}
\label{Fig.3}
\end{figure}

\begin{figure}
\caption{Same as Fig.~\protect\ref{Fig.2} for the ($T=0,\,S=1$) spin-orbit
potential.}
\label{Fig.4}
\end{figure}

\begin{figure}
\caption{Same as Fig.~\protect\ref{Fig.2} for the ($T=1,\,S=1$) tensor
potential.}
\label{Fig.5}
\end{figure}

\begin{references}

\bibitem{TS73}
R. de Tourreil and D.W.L. Sprung, {\em Nucl. Phys.} {\bf A201}, 193 (1973).

\bibitem{TRS75}
R. de Tourreil, B. Rouben, and D.W.L. Sprung,
{\em Nucl. Phys.} {\bf A242}, 445 (1975).

\bibitem{Lac+80}
M. Lacombe, B. Loiseau, J.M. Richard, R. Vinh Mau, J. C\^ot\'e,
P. Pires, and R. de Tourreil, {\em Phys. Rev. C} {\bf 21}, 861 (1980).

\bibitem{MHE87}
R. Machleidt, K. Holinde, and C. Elster, {\em Phys. Rep.}
{\bf 149}, 1 (1987).

\bibitem{WSA84}
R.B. Wiringa, R.A. Smith, and T.L. Ainsworth, {\em Phys. Rev. C}
{\bf 29}, 1207 (1984).

\bibitem{Nij78}
M.M. Nagels, T.A. Rijken, and J.J. de Swart,
{\em Phys. Rev. D} {\bf 17}, 768 (1978).

\bibitem{ER82}
T.E.O. Ericson and M. Rosa-Clot, {\em Phys. Lett.} {\bf 110B}, 193 (1982);
{\em Nucl. Phys.} {\bf A405}, 497 (1983).

\bibitem{BER89}
J.L. Ballot, A. Eir\'o, and M.R. Robilotta,
{\em Phys. Rev. C} {\bf 40}, 1459 (1989).

\bibitem{BR92}
J.L. Ballot and M.R. Robilotta, {\em Phys. Rev. C} {\bf 45}, 986 (1992).

\bibitem{BR95}
J.L. Ballot and M.R. Robilotta, {\em J. Phys. G}, {\bf 20}, 1599 (1994).

\bibitem{Wei66}
S. Weinberg, {\em Phys. Rev. Lett.} {\bf 17}, 616 (1966).

\bibitem{Tom67}
H.S. Mani, Y. Tomozawa, and Y.P. Yao, {\em Phys. Rev. Lett.}
{\bf 18}, 1084 (1967).

\bibitem{CWZ69}
S. Coleman, J. Wess, and B. Zumino, {\em Phys. Rev.} {\bf 177}, 2239 (1969);
C.G. Callan, S. Coleman, J. Wess, and B. Zumino, {\em Phys. Rev.} {\bf 177},
2247 (1969).

\bibitem{CF86}
S.A. Coon and J.L. Friar, {\em Phys. Rev. C} {\bf 34}, 1060 (1986).

\bibitem{BD71}
G.E. Brown and J.W. Durso, {\em Phys. Lett.} {\bf 35B}, 120 (1971).

\bibitem{CDR72}
M. Chemtob, J.W. Durso, and D.O. Riska {\em Nucl. Phys} {\bf B38}, 141 (1972).

\bibitem{PL70}
M.H. Partovi and E.L. Lomon, {\em Phys. Rev.} {\bf D2}, 1999 (1970).

\bibitem{PL72}
F. Partovi and E.L. Lomon, {\em Phys. Rev. D} {\bf 5}, 1192 (1972).

\bibitem{ZT82}
M.J. Zuilhof and J.A. Tjon, {\em Phys. Rev. C} {\bf 24},
736 (1981); {\bf 26}, 1277 (1982).

\bibitem{OVK92}
C. Ord\'o\~nez and U. Van Kolk, {\em Phys. Lett.} {\bf B291}, 459 (1992).

\bibitem{Wei90}
S. Weinberg, {\em Phys. Lett. B} {\bf 251}, 288 (1990); {\em Nucl. Phys.}
{\bf B363}, 3 (1991).

\bibitem{CPS92}
L.S. Celenza, A. Pantziris and C.M. Shakin, {\em Phys. Rev. C}
{\bf 46}, 2213 (1992).

\bibitem{Bir94}
M.C. Birse, {\em Phys. Rev. C} {\bf 49}, 2212 (1994).

\bibitem{FC94}
J.L. Friar and S.A. Coon, {\em Phys. Rev. C} {\bf 49}, 1272 (1994).

\bibitem{RR94}
C.A. da Rocha and M.R. Robilotta, {\em Phys. Rev. C}, {\bf 49}, 1818 (1994).

\bibitem{SB51}
E.E. Salpeter e H.A. Bethe, {\em Phys. Rev.} {\bf 84}, 1232 (1951).

\bibitem{BW53}
K.A. Brueckner e K.M. Watson,
{\em Phys. Rev.} {\bf 90}, 699; {\bf 92}, 1023 (1953).

\bibitem{BS66}
R. Blankenbecler e R. Sugar, {\em Phys. Rev.} {\bf 142}, 1051, (1966).

\bibitem{Pup94}
J.C. Pupin, {\em M.Sc. Thesis}, Instituto de F\'{\i}sica Te\'orica,
Universidade Estadual Paulista, S\~ao Paulo, 1994 (unpublished).

\bibitem{ORK94}
C. Ord\'o\~nez, L. Ray, and U. Van Kolck,
{\em Phys. Rev. Lett.} {\bf 72}, 1982 (1994).

\end{references}
\end{document}